\documentclass{webofc}
\usepackage[varg]{txfonts}   % Web of Conferences font
\newcommand{\bs}{\boldsymbol}

\newcommand{\eq}[1]{\begin{align} #1 \end{align}}

\begin{document}
\title{van der Waals Interactions and Hadron Resonance Gas: Role of resonance widths modeling on conserved charges fluctuations\thanks{Presented at \emph{Strangeness in Quark Matter 2017} conference, Utrecht, Netherlands, 10-15 July 2017}}

\author{
\firstname{Volodymyr} \lastname{Vovchenko}\inst{1,2}\fnsep\thanks{\email{vovchenko@fias.uni-frankfurt.de}} \and
\firstname{Paolo} \lastname{Alba}\inst{2} \and
\firstname{Mark I.} \lastname{Gorenstein}\inst{2,3} \and
\firstname{Horst} \lastname{Stoecker}\inst{1,2,4}
}

\institute{Institut f\"ur Theoretische Physik,
Goethe Universit\"at Frankfurt, %Max-von-Laue-Str. 1, 
D-60438 Frankfurt am Main, Germany
\and
           Frankfurt Institute for Advanced Studies, Giersch Science Center, %Ruth-Moufang-Str. 1, 
           D-60438 Frankfurt am Main, Germany
\and
           Bogolyubov Institute for Theoretical Physics, 03680 Kiev, Ukraine
\and
           GSI Helmholtzzentrum f\"ur Schwerionenforschung GmbH, %Planckstr. 1, 
           D-64291 Darmstadt, Germany
          }

\abstract{%
The quantum van der Waals (QvdW) extension of the ideal hadron resonance gas (HRG) model which includes the attractive and repulsive interactions between baryons -- the QvdW-HRG model -- is applied to study the behavior of the baryon number related susceptibilities in the crossover temperature region.
Inclusion of the QvdW interactions leads to a qualitatively different behavior of susceptibilities, in many cases resembling lattice QCD simulations.
It is shown that for some observables, in particular for $\chi_{11}^{BQ} / \chi_2^B$, effects of the QvdW interactions essentially cancel out.
It is found that the inclusion of the finite resonance widths leads to an improved description of $\chi_2^B$, but it also leads to a worse description of $\chi_{11}^{BQ} / \chi_2^B$, as compared to the lattice data.
On the other hand, inclusion of the extra, unconfirmed baryons into the hadron list leads to a simultaneous improvement in the description of both observables.
}
\maketitle
\section{Introduction}
\label{sec:intro}
Lattice QCD simulations provide the equation of state of strongly interacting matter at zero net baryon density~\cite{Borsanyi:2013bia,Bazavov:2014pvz}.
A smooth crossover-type transition between hadronic and partonic matter is observed~\cite{Aoki:2006we}.
A common model for the hadronic phase -- the ideal hadron resonance gas (IHRG) model -- successfully describes many lattice observables at lower temperatures, $T \sim 100-150$~MeV.
Agreement of the IHRG model with lattice data rapidly breaks down at $T \simeq 150-160$~MeV for fluctuations and correlations of conserved charges~\cite{Bazavov:2013dta}. 
Sometimes, this breakdown was interpreted as a signal for deconfinement~\cite{Bazavov:2013dta,Bazavov:2013uja}.

On the other hand, it has recently been shown that the onset of deviations of IHRG from the lattice data can well be explained by the QvdW-type interactions between baryons~\cite{Vovchenko:2016rkn}.
In this work we assess simultaneous effects of QvdW interactions, the modeling of the finite widths of the resonances, and also the HRG hadron list.
The importance of the potentially missing hadron states in the hadron list was pointed out in several recent publications, in the context of the IHRG model~\cite{Majumder:2010ik,Bazavov:2014xya,Bazavov:2017dus,Alba:2017mqu} and also in the context of the excluded volume (EV) HRG model~\cite{Alba:2017bbr}, while the effects due to the finite widths of the resonances were barely discussed at all.
None of the two effects were studied simultaneously with the presence of the QvdW interactions.

\section{Model}
\label{sec:model}

{\bf QvdW extension.} The Quantum van der Waals extension of the HRG model of Ref.~\cite{Vovchenko:2016rkn} -- the QvdW-HRG model -- is based on the following assumptions:

1. QvdW interactions exist between all pairs of baryons and between all pairs of antibaryons. The QvdW parameters $a$ and $b$ for all (anti)baryons taken to be equal to those of nucleons, as obtained from the fit to the ground state of nuclear matter~\cite{Vovchenko:2015vxa}: $a \simeq 329$~MeV fm$^3$ and $b \simeq 3.42$~fm$^3$.

2. The baryon-antibaryon, meson-meson, and meson-(anti)baryon QvdW interactions are neglected.
Note that the model still contains, by construction, the meson-related hadronic interactions that lead to the formation of resonances.

The QvdW-HRG model contains basic nuclear matter physics, in contrast to IHRG, and it yields the liquid-gas first-order phase transition in the symmetric nuclear matter with a critical point located at $T_c \simeq 19.7$~MeV and $\mu_c \simeq 908$~MeV ($n_c \simeq 0.07$~fm$^3 = 0.45\,n_0$).

The QvdW-HRG consists of three
sub-systems: Ideal gas of mesons, QvdW gas of baryons, and QvdW gas of antibaryons. The total pressure reads
$p(T,\bs \mu) = p_M(T,\bs \mu) + p_B(T,\bs \mu) + p_{\bar{B}}(T,\bs \mu)$
with
\begin{align}
\label{eq:Ps}
p_M(T,\bs \mu) =
\sum_{j \in M} p_{j}^{\rm id} (T, \mu_j), \qquad
p_{B(\bar{B})}(T,\bs \mu) =
\sum_{j \in B(\bar{B})} p_{j}^{\rm id} (T, \mu_j^{B(\bar{B})*}) - a\,n_{B(\bar{B})}^2 ,
\end{align}
where $M$ stands for mesons, $B(\bar{B})$ for (anti)baryons, 
$\bs \mu=(\mu_B,\mu_S,\mu_Q)$ are the chemical potentials 
for net baryon number $B$, strangeness $S$, and electric charge $Q$,
$\mu_j^{B(\bar{B})*} = \mu_j - b\,p_{B(\bar{B})} - a\,b\,n_{B(\bar{B})}^2 + 2\,a\,n_{B(\bar{B})}$,
$\mu_j = B_j \, \mu_B + S_j \, \mu_S + Q_j \, \mu_Q$ is the chemical potential for hadron species $j$, with $B_j$, $S_j$, and $Q_j$ being its corresponding quantum numbers.
$n_B$ and $n_{\bar{B}}$ are total densities of baryons and antibaryons.

The calculation of mesonic pressure $p_M(T,\bs \mu)$ is straightforward.
The shifted chemical potentials $\mu_j^{B(\bar{B})*}$ of (anti)baryons depend explicitly on (anti)baryon pressure $p_{B(\bar{B})}$ and on total (anti)baryon density $n_{B(\bar{B})}$. By taking the derivatives of $p_{B(\bar{B})}$ with respect to the baryochemical potential one obtains additional equation for the total (anti)baryon densities: $n_{B(\bar{B})} = (1 - b \, n_{B(\bar{B})})  \sum_{j \in B(\bar{B})} n_{j}^{\rm id} (T, \mu_j^{B(\bar{B})*})~.$

At given $T$ and $\bs \mu$, the above equations 
are solved numerically, yielding $p_{B(\bar{B})}(T,\bs \mu)$ and $n_{B(\bar{B})}(T,\bs \mu)$.
The entropy density is $s = (\partial p / \partial T)_{\mu}$, and the energy density is obtained from the Gibbs relation.

\vskip5pt

\noindent{\bf Finite widths of the resonances.} 
The ideal Fermi or Bose gas pressures $p_{i}^{\rm id} (T, \mu_i)$ in Eq.~\eqref{eq:Ps} contain the additional integration over hadron's mass:
\eq{\label{eq:p}
p_{i}^{\rm id} (T, \mu_i) =  \frac{d_i}{6 \pi^2} \int_{m_i^{\rm min}}^{m_i + 2\Gamma_i} d m \, \rho_i(m)  \, \int dk \, \frac{k^4}{\sqrt{m^2 + k^2}} \, \left[\exp\left( \frac{\sqrt{m^2+k^2} - \mu_i}{T} \right) \pm 1 \right]^{-1}.
}
For $n_{j}^{\rm id} (T, \mu_i)$ the expressions are analogous to~\eqref{eq:p}.
The function $\rho_i(m)$ is the properly normalized mass distribution for hadron type $i$.
For stable hadrons, or whenever the zero width approximation is applied, one has $\rho_i(m) = \delta(m - m_i)$.
The finite widths of the resonances are taken into account in a simplified way, by the integration over their relativistic Breit-Wigner shapes~(see, e.g., Refs.~\cite{Becattini:1995if,Wheaton:2004qb}):
\eq{\label{eq:rho}
\rho_i(m) = \frac{2 \, m \, m_i \, \Gamma_i}{(m^2-m_i^2)^2 + m_i^2 \Gamma_i^2} \bigg/ \left(\int_{m_i^{\rm min}}^{m_i + 2\Gamma_i} d \tilde{m} \,
\frac{2 \, \tilde{m} \, m_i \, \Gamma_i}{(\tilde{m}^2-m_i^2)^2 + m_i^2 \Gamma_i^2}\right),
}
where $m^{\rm min}_i = \operatorname{max}(m_i - 2 \Gamma_i, m^{\rm thr}_i)$ with $m^{\rm thr}_i$ the minimum decay threshold mass for resonance $i$.

\vskip5pt

\noindent{\bf Hadron list.} In the standard scenario, the hadron list includes all established~(3- and 4-star) strange and non-strange hadrons which are listed in the Particle Data Tables~\cite{Agashe:2014kda}. This list is denoted as PDG and it contains about 380 different hadron species.
In addition, we also consider an extended PDG-based list, which also includes unconfirmed hadron states. This PDG+ list contains about 580 hadron species.
Using the PDG and PDG+ lists we test the sensitivity of the results to the input hadron list.

\section{Results}

\begin{figure*}[t]
\begin{center}
\includegraphics[width=0.325\textwidth]{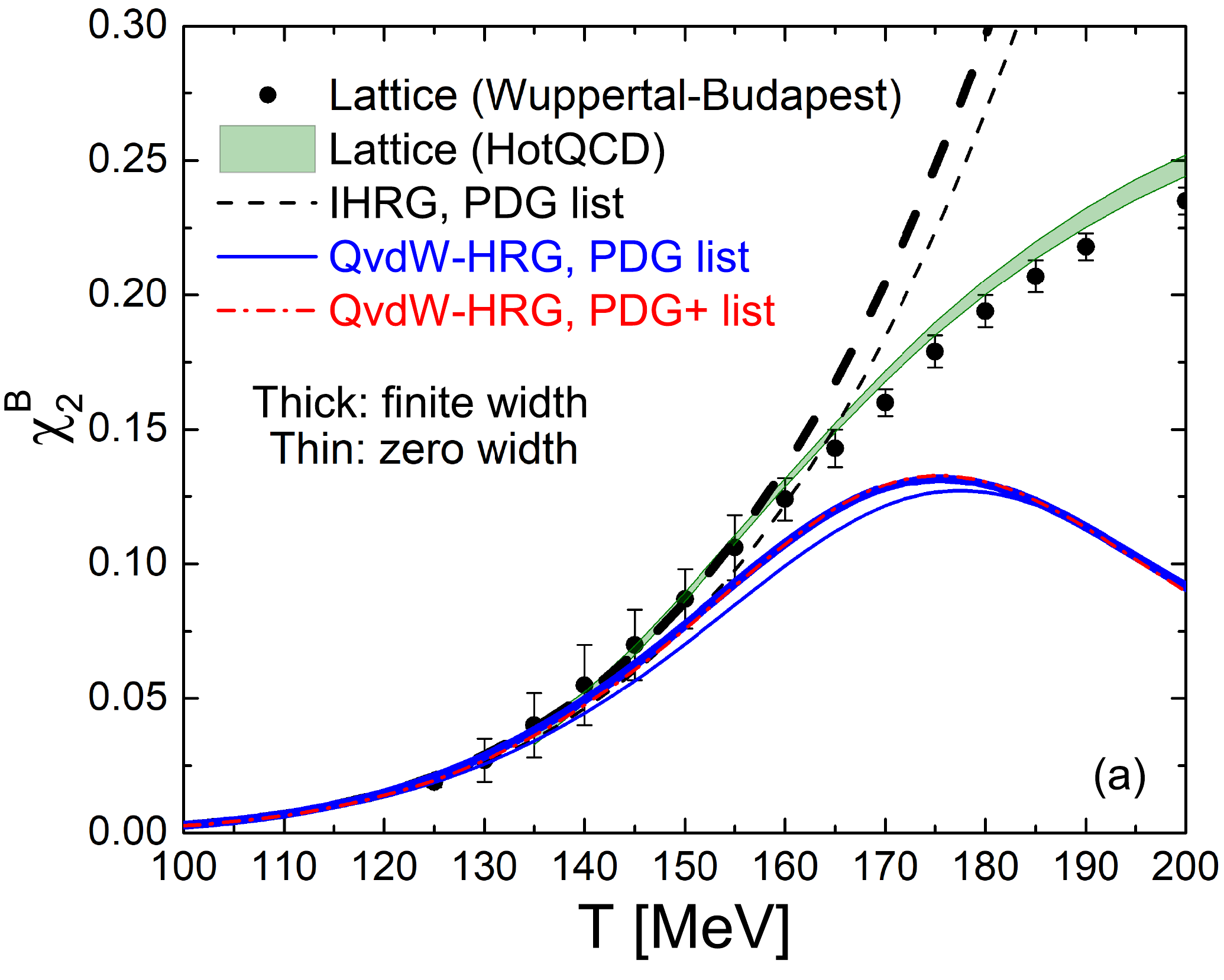}
\includegraphics[width=0.325\textwidth]{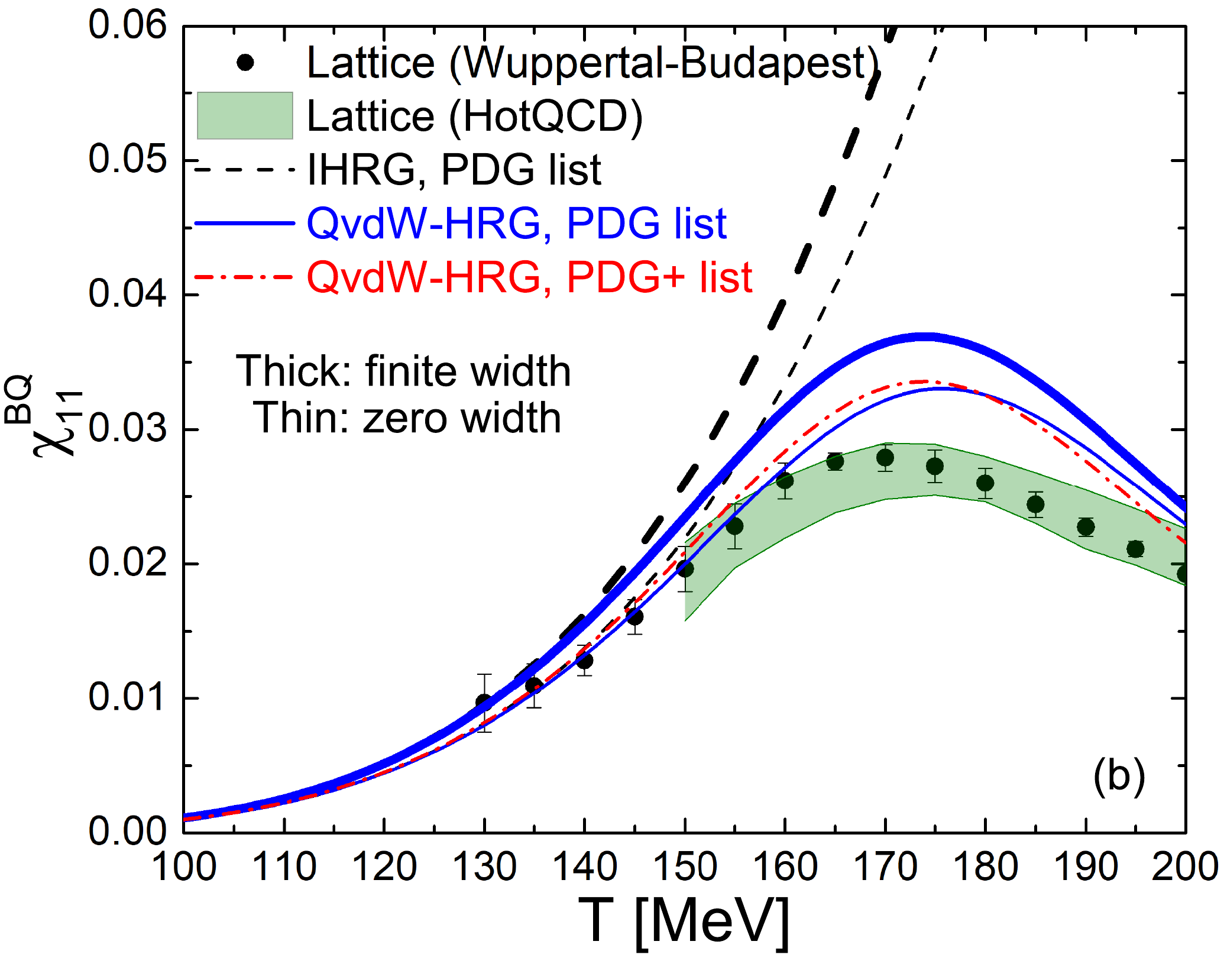}
\includegraphics[width=0.325\textwidth]{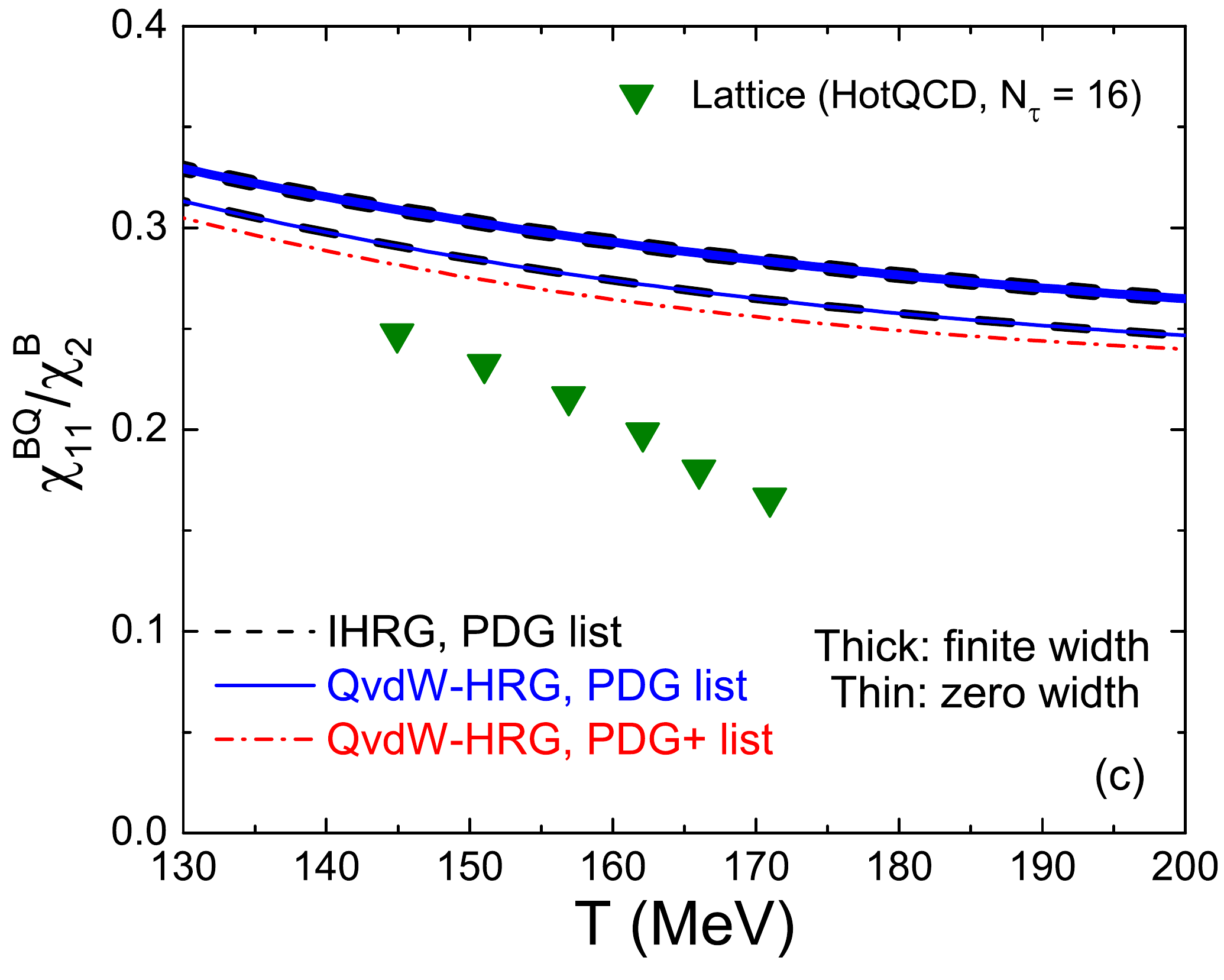}
\caption{The temperature dependence of (a) net baryon number susceptibility $\chi_2^B$, (b) baryon-electric charge correlator $\chi_{11}^{BQ}$, and (c) $\chi_{11}^{BQ} / \chi_2^B$,
calculated within IHRG (dashed black lines) and QvdW-HRG (solid blue lines) with PDG hadron list. Calculations with finite (zero) widths of the resonances are depicted by the thick (thin) lines.
QvdW-HRG calculations with PDG+ hadron list are exhibited by the dash-dotted red lines.
Lattice QCD results of the Wuppertal-Budapest~\cite{Borsanyi:2011sw,Borsanyi:2015axp} and HotQCD~\cite{Bazavov:2012jq,Bazavov:2017dus,Karsch:2017zzw} collaborations are shown, respectively, by black symbols and green bands/symbols.
}
\label{fig1}
\end{center}
\end{figure*}

We consider temperature dependence of the conserved charges susceptibilities at $\mu = 0$, defined as
\eq{
\label{eq:BSQ}
\chi_{lmn}^{BSQ}~=~\frac{\partial^{l+m+n}p/T^4}{\partial(\mu_B/T)^l \,\partial(\mu_S/T)^m \,\partial(\mu_Q/T)^n}~\,.
}
We focus on the net baryon susceptibility $\chi_2^B$, baryon-electric charge correlator $\chi_{11}^{BQ}$, and the ratio $\chi_{11}^{BQ}$ / $\chi_2^B$.
The temperature dependences of these quantities are shown in Fig.~\ref{fig1}, along with the lattice data.

The inclusion of QvdW interactions between baryons leads to a qualitatively different behavior of $\chi_2^B$ and $\chi_{11}^{BQ}$ at high temperatures, as seen from comparison between the dashed black lines (IHRG) and the solid blue lines (QvdW-HRG) in Fig.~\ref{fig1}. 
Some qualitative features seen in lattice simulations, such as the inflection point in the temperature dependence of $\chi_2^B$ and the peak in $\chi_{11}^{BQ}$ are reproduced by the QvdW-HRG model, as reported previously in Ref.~\cite{Vovchenko:2016rkn}.

It is interesting that the ratio $\chi_{11}^{BQ}$ / $\chi_2^B$~(Fig.~\ref{fig1}c) is virtually unaffected by the QvdW interactions. 
This result can be proved analytically for the Boltzmann approximation. 
In this case the ratio $n_i / n_i^{\rm id}$ between the density $n_i$ of the baryon species $i$ in the QvdW-HRG model and the corresponding ideal gas density $n_i^{\rm id}$ is the same for all baryons and is a function of the total density of baryons $n_B$ only. At $\mu_B = 0$ one has $n_{\bar{B}} = n_B$, therefore the modification factor is the same for baryons and antibaryons.
Therefore, the effects of QvdW interactions between baryons cancel out in the ratio $\chi_{11}^{BQ}$ / $\chi_2^B$ at $\mu_B = 0$:
\eq{
\left(\chi_{11}^{BQ} / \chi_2^B\right)_{\rm QvdW} = \frac{\sum_{i \in B} Q_i \, n_i^{\rm id}}{\sum_{i \in B} n_i^{\rm id}} = \left(\chi_{11}^{BQ} / \chi_2^B\right)_{\rm id}~.
}
This explains the results shown in Fig.~\ref{fig1}c. Note that, in general, there is no such cancellation if the QvdW interaction parameters would be assumed to be different for different baryon-baryon pairs~\cite{Alba:2017bbr}.

All three observables, including the ratio $\chi_{11}^{BQ} / \chi_2^B$, are sensitive to the modeling of the finite widths of the resonances, in particular the $\Delta$ and $N^*$ resonances.
Applying the prescription given by Eqs.~\eqref{eq:p} and \eqref{eq:rho} one obtains an improved description of $\chi_2^B$, but also a worse description of $\chi_{11}^{BQ}$ and $\chi_{11}^{BQ} / \chi_2^B$, within both the IHRG and QvdW-HRG models.
The influence of the finite resonance widths is also the likely source of the discrepancy between the IHRG and QvdW-HRG results for $\chi_{11}^{BQ} / \chi_2^B$ reported in Ref.~\cite{Karsch:2017zzw}.
Present results imply a necessity for a more involved modeling of the resonances in a HRG.
One possibility 
is to use the S-matrix approach~\cite{Lo:2017lym}.

The description of all the considered observables is improved when an extended PDG+ hadron list, which contains additional baryons, is used. The improvement is rather modest, and the lattice data for the $\chi_{11}^{BQ} / \chi_2^B$ ratio are still not described well by all the considered models.
A notably improved description can be obtained by using the quark model states~\cite{Karsch:2017zzw}.
Another interesting possibility, also presented at this conference, are the in-medium mass modifications for the negative-parity states~\cite{Aarts:2017rrl,Aarts:2017iai}.
It would be interesting to compare and combine these modifications with the QvdW approach.

To summarize,
the effects of quantum van der Waals interactions in the HRG model on the baryon number susceptibilities are studied simultaneously with the effects of finite resonance widths and input hadron list.
QvdW interactions lead to a qualitatively different behavior of $\chi_2^B$ and $\chi_{11}^{BQ}$, some features resembling the lattice data, but they cancel out in the ratio $\chi_{11}^{BQ} /\chi_2^B$.
The inclusion of finite resonance widths via the relativistic Breit-Wigner distribution improves the description for $\chi_2^B$, but also leads to a worse agreement for $\chi_{11}^{BQ}$ and $\chi_{11}^{BQ} /\chi_2^B$.
Inclusion of the extra, unconfirmed hadron states from PDG appears to improve slightly the agreement for all the observables considered.

\textit{\textbf{Acknowledgments.}} We thank F. Karsch for useful discussions regarding $\chi_{11}^{BQ} /\chi_2^B$ and for providing the lattice data for this quantity. We acknowledge support from HIC for FAIR, HGS-HIRe, Judah M. Eisenberg Laureatus Chair at Goethe University, and National Academy of Sciences of Ukraine.

\end{document}